# Highlights

## Explainable-AI Powered stock price prediction using time series transformers: A Case Study on BIST100

Şükrü Selim Çalık, Andaç Akyüz, Zeynep Hilal Kilimci, Kerem Çolak

- Transformer-based architectures are employed for stock price forecasting of the top five high-volume banks listed in the BIST100 index, as well as the XBANK and XU100 indices.
- The DLinear model outperforms all other models across multiple performance metrics, demonstrating superior forecasting capability.
- Feature sets are enriched with a wide range of technical indicators to enhance predictive accuracy.
- Explainable AI techniques, including SHAP and LIME, are integrated to provide transparency and interpretability of model outputs.
- The study illustrates how advanced forecasting models can be aligned with financial literacy objectives by making complex tools more understandable and actionable.
- Empirical findings highlight the potential of AI-driven methods to support more informed investment decisions and encourage active participation in capital markets.
- The proposed approach offers a replicable framework for combining artificial intelligence and financial literacy efforts, particularly in emerging market contexts.

# Explainable-AI Powered stock price prediction using time series transformers: A Case Study on BIST100

Şükrü Selim Çalık[a,b], Andaç Akyüz[b], Zeynep Hilal Kilimci[b] and Kerem Çolak[c]

[a]*Maviay Consultancy Company, Kocaeli University Technopark, Kocaeli, Türkiye*
[b]*Department of Information Systems Engineering, Kocaeli University, Kocaeli, Türkiye*
[c]*Hereke Asım Kocabıyık Vocational School, Kocaeli University, Türkiye*

ARTICLE INFO

ABSTRACT

Financial literacy is increasingly dependent on the ability to interpret complex financial data and utilize advanced forecasting tools. In this context, this study proposes a novel approach that combines transformer-based time series models with explainable artificial intelligence (XAI) to enhance the interpretability and accuracy of stock price predictions. The analysis focuses on the daily stock prices of the five highest-volume banks listed in the BIST100 index, along with XBANK and XU100 indices, covering the period from January 2015 to March 2025. Models including DLinear, LTSNet, Vanilla Transformer, and Time Series Transformer are employed, with input features enriched by technical indicators. SHAP and LIME techniques are used to provide transparency into the influence of individual features on model outputs. The results demonstrate the strong predictive capabilities of transformer models and highlight the potential of interpretable machine learning to empower individuals in making informed investment decisions and actively engaging in financial markets.

## 1. Introduction

The stock market is without doubt one of the most important markets, keeping the pulse of both the global economy and national economies. The functioning of this market and the formation of its prices are directly related to complex variables. Economic data, investor psychology and external news flows shape the dynamic structure of stock market prices [33]. This inherent complexity makes it challenging to predict prices and make investment decisions. The prevailing stock price at a given moment is determined by the most recent price that has been accepted by both the buyer and the seller. It can thus be concluded that the stock price is subject to regulation by the forces of supply and demand. The total quantity of shares available in a company corresponds to supply; however, demand is influenced by a multitude of factors, which can be variable in their predictability. In order to achieve success in the realm of investment, it is imperative for investors to make judicious decisions regarding the acquisition or retention of their financial assets. However, the stock market is characterised by high volatility, non-linear dynamics and a multitude of uncertainties influenced by interconnected economic and political factors on a global scale. The factors that may be considered include, but are not limited to, inflation, interest rates, news, oil prices, war, terrorism, and political instability. Nevertheless, uncertainty is recognised as a pivotal factor in market dynamics [11]. The generation of accurate methods for predicting stock price changes has been a long-standing objective of the financial and technological communities. In this context, the prediction of future movement patterns in stock prices has been a subject of extensive research in the academic literature. While there are proponents of the efficient market hypothesis who believe that it is impossible to predict stock prices, there are also propositions that demonstrate that, if properly formulated and modelled, stock price forecasts can be made with a fairly high level of accuracy. The construction of robust statistical, econometric and machine learning models is possible through the careful selection of variables and appropriate functional forms or forecasting models [14]. Nevertheless, financial literacy remains a pivotal aspect for individual investors when undertaking investment decisions within the stock market. Financial literacy is defined as the set of skills and knowledge that enables individuals to understand financial products and markets, and to assess risks. Nevertheless, it is important to note that there may be instances where the necessary data and analytical tools are not readily available, particularly for investors with limited financial resources. In the contemporary context, technological tools such as machine learning have emerged as a potent solution to enhance financial literacy and facilitate informed investment decision-making [9]. The practice of stock market price forecasting, which is supported by a range of tools including traditional financial methods, time series analyses, econometric models and statistical analyses, is now being advanced further through the use of machine learning-based models. In contemporary research, sophisticated methodologies such as artificial neural networks, deep learning techniques and genetic algorithms are employed in intricate processes, including market forecasting and behavioural analysis [27]. Consequently, forecasting models supported by machine learning overcome the limitations of traditional methods and offer the opportunity to explain more flexible and complex market dynamics. Despite the fundamental role that conventional techniques play in the realm of stock market forecasting, it is imperative to acknowledge their potential limitations when confronted with elevated levels of uncertainty. Conversely, machine learning, leveraging its substantial computational capacity and aptitude for learning from historical data, has the potential to generate more reliable and effective forecasting models, particularly in market structures characterised by non-linear and complex relationships [18]. Consequently, from the perspective of enhancing financial literacy and empowering individual investors, the integration of machine learning with conventional methodologies can be regarded as a strategic initiative. As demonstrated in the extant literature, artificial intelligence (AI), employing machine learning and deep learning methodologies, has been shown to yield more efficient outcomes in non-linear forecasting applications, attributable to its sophisticated computational capacities. This is due to the fact that AI is capable of acquiring and retaining information in a manner analogous to the human brain [2]. In summary, the system has the capacity to select the information to be used for prediction from past data. The methods employed within the field of artificial intelligence differ in terms of memory structures, the number of layers that can be utilised, and the areas in which they are employed [29]. In this context, a machine learning–based guidance system for individual investors could provide a more informed approach to investment decisions and offer greater protection against potential financial threats.

*Corresponding author.
✉ sselimcalik1@gmail.com (S. Çalık); akyuzandac@gmail.com (A. Akyüz); zeynep.kilimci@kocaeli.edu.tr (Z.H. Kilimci); kerem.colak@kocaeli.edu.tr (K. Çolak)
ORCID(s):

The field of Explainable Artificial Intelligence (XAI) is distinct from other AI algorithms in its objective to enhance the transparency and comprehensibility of decision-making processes within artificial intelligence and machine learning models. The utilisation of XAI has been demonstrated to enhance user confidence in the aforementioned algorithms by elucidating the internal mechanisms of the models and the rationale underpinning the prediction outcomes. The utilisation of XAI in the domains of stock market and financial market predictions facilitates enhanced comprehension of the outcomes derived from the model by investors, thereby empowering them to make more informed investment decisions. By providing greater transparency regarding predicted market movements, XAI prevents investors from relying solely on a 'black box' and thus increases financial literacy. This feature of XAI is highly valuable for both individual investors and financial analysts, as it enables the clear assessment of predicted risks and opportunities, thereby facilitating the development of informed strategies [8], [10].

To contribute to the growing body of research on financial literacy and its impact on investor behavior, this study adopts a novel approach by integrating cutting-edge transformer-based time series models with explainable artificial intelligence (XAI) techniques to forecast stock prices in the Turkish banking sector. Unlike traditional methods, transformer architectures—DLinear, LTSNet, Vanilla Transformer and Time Series Transformer(TST)—are employed to capture the complex, nonlinear, and temporal dynamics of financial markets with heightened precision. To enhance the relevance of the predictions for real-world decision-making, the models are enriched with a comprehensive set of technical indicators and subjected to interpretability analyses using SHAP and LIME. This integration of high-performing transformer models and XAI not only advances predictive accuracy but also renders complex model outputs more transparent and accessible for non-expert users. In doing so, the study directly supports the development of practical financial literacy by empowering individuals with interpretable and data-driven tools for market participation, particularly in emerging market economies.

The key contributions of this study are as follows:

1. Introduction of advanced transformer-based models to the task of stock price forecasting in an emerging market context, providing new insights into their effectiveness relative to one another.
2. Integration of explainable AI (XAI) techniques (SHAP and LIME) into financial forecasting, enhancing model transparency and interpretability for both academic and practical applications.
3. Support for financial literacy by demonstrating how complex predictive tools can be made comprehensible and useful for investors, thereby promoting informed decision-making.
4. Use of enriched feature sets including a wide array of technical indicators to improve the forecasting power of transformer models.
5. Provision of a replicable, policy-relevant framework for combining AI technologies with financial literacy goals, particularly suited to the needs of developing and emerging markets.

The remainder of this paper is organized as follows. The Related Work section reviews recent literature on stock price forecasting in financial applications. The Proposed Framework section details the theoretical background, application framework, data collection process, the transformer-based forecasting models and XAI methods implemented in this study. The Experiment Results section presents the forecasting results. The Discussion section interprets the findings with a focus on model performance and the implications for financial literacy. Finally, the Conclusion section summarizes the main contributions of the study and outlines potential directions for future research.

## 2. Related work

In the field of research concerning the financial implications of artificial intelligence, The study [19] has been a significant contribution. Examining both the impact of AI in the financial sector and its relationship with financial literacy, the author's primary focus is on how financial technologies, otherwise termed fin-tech, are effecting the transformation of financial services in developing countries. The study indicates that artificial intelligence (AI) and fin-tech companies are instrumental in enhancing accessibility to financial services and integrating them into the daily lives of citizens. However, in countries such as India, where financial literacy levels are low, this results in a significant loss of opportunity. The increased investment in financial technology is indicative of its rapid growth and the activation of decision-making mechanisms that can engender enhanced financial well-being. The author posits that the effective utilisation of these technologies is contingent upon a foundation of financial literacy, underscoring the critical importance of enhancing financial awareness and comprehension. Concurrently, advancements in these technologies are anticipated to exert a favourable influence on financial literacy.

In the study [15], Leung et al., introduced a portfolio recommendation system to the literature, offering a profitable stock portfolio and stock analytics using machine learning and big data analytics. The effectiveness of the system was evaluated through a two-part user assessment and performance evaluation via backtesting. The findings demonstrate that a portfolio recommendation system founded upon machine learning and big data analytics is capable of effectively meeting the expectations of the majority of users and enhancing their financial knowledge. The study under discussion sheds light on the potential of machine learning and big data analytics in the financial sector.

Given the pivotal role of stock trading within the financial world, the study [24] conducted a study on stock market prediction with the objective of ascertaining the future value of a stock. In the present study, the utilisation of machine learning techniques in the domain of stock prediction is elucidated. The majority of stock brokers utilise a combination of technical and fundamental analysis, as well as time series methods, employing Python programming in the process. The proposed machine learning approach ensures accuracy by learning from existing stock data. Support Vector Machines (SVM) are employed to predict stock prices for both large and small market-cap stocks using large data sets. Support vector machines (SVMs) have been developed to address the issue of overfitting by creating multiple models capable of predicting daily market trends. The numerical outcomes demonstrate that these models exhibit high efficiency. The utilisation of trained predictors in the creation of practical trading models has been demonstrated to yield higher returns in comparison to established benchmarks.

The study [26] emphasised the importance of "Big Data" and "Parallel Processing" methods in problem-solving processes was highlighted. The research conducted predicted IBM stock prices traded on the New York Stock Exchange. The researchers developed a model using three different deep learning frameworks—LSTM, GRU, and BLSTM—for IBM stock predictions, relying on the assumption that deep learning algorithms implemented on multicore computing devices like GPUs yield significant success in solving real-world problems. A series of experiments were conducted utilising data from the New York Stock Exchange from 1968 to 2018. The findings indicated that the BLSTM model demonstrated a directional accuracy rate of 63.54% when prompted with the previous five days' transaction data. Furthermore, a 13.47% profit was realised after a 10-day period. Researchers have proposed the optimisation of algorithms and adjustment of parameters with a view to enhancing system efficiency and facilitating more accurate predictions.

In a separate study on forecasting stock indices and prices, the study [18] presented two approaches. Initially, the Feed-Forward Neural Network was employed with back-propagation training, achieving an average accuracy of 97.66%. Nevertheless, this model necessitates substantial training data and epochs, and is susceptible to overfitting issues. Secondly, the Convolutional Neural Network (CNN) model was determined to be more effective in analysing time series data and made predictions by using grayscale 2-D histograms. The dataset was segmented into 15 sections, with each segment being fed into the CNN model, resulting in high accuracy. This model required less training data and time compared to the previous model and achieved an average accuracy of 98.92%. It is evident that both methods have their respective advantages and disadvantages. However, it is important to note that they are capable of making accurate predictions for stock markets. This, in turn, aids analysts in forecasting future market patterns for companies and economies.

The study [12] conducted a study to ascertain the validity of the "semi-strong market efficiency" hypothesis in relation to Borsa Istanbul. The directions of 26 highly liquid stocks' 5-minute closing price movements were predicted using machine learning algorithms, classified as up, down, or stable. With regard to the effective utilisation of Borsa Istanbul's "data analytics" set, it was determined that the data does not fully reflect prices, and market efficiency has not been achieved. The application of diverse classifiers yielded the identification of nine stocks as being susceptible to prediction. However, it was observed that the predictability of these stocks diminished with the prolongation of the timeframe over which data was collected. In the context of economic gain analysis, a number of stocks have demonstrated superior performance when utilising short-selling strategies. The study's findings indicate that data from Borsa Istanbul is not being used effectively by market participants.

In a study [1] conducted for the BIST Banking sector, the authors analysed price forecasting methods for the BIST Banks Index. The study utilised daily closing data from 27 December 1996 to 31 August 2023. The research compares the traditional ARIMA model with AI-based deep learning models, the Financial Market Model (FPM), and the Convolutional Neural Network Model (CNNM). While the conventional analysis method ARIMA yielded superior outcomes in price forecasting when compared with AI models, a comparative analysis of AI models revealed that CNNM exhibited superior prediction accuracy when compared with FPM. The study demonstrates that artificial intelligence (AI) and deep learning techniques offer a more effective alternative to traditional models in forecasting financial asset prices. Financial literacy is of critical importance in the employment of both traditional ARIMA and AI methods. The research is subject to certain limitations, including the length of the prediction times and the high hardware requirements. It is suggested that future studies should explore different price forecasting models and financial indices in order to enhance the efficiency of AI.

In another comparative analysis [7], authors examine the impact of artificial intelligence (AI) on financial analysis and forecasting, with a particular focus on the banking sector. The capacity of AI to process voluminous datasets and enhance prediction accuracy is considered to be pivotal for the enhancement of financial decision-making processes.

In the analysis of financial data, both conventional statistical methodologies (ARIMA models) and machine learning algorithms (Gradient Boosting Machines and Random Forests) were utilised. The integration of artificial intelligence (AI) has been demonstrated to enhance prediction accuracy by 30% and risk assessment accuracy by 20%. It is noteworthy that Gradient Boosting Machines exhibited superiority in the identification of investment portfolio risks.

The objective of this study is to demonstrate that an individual with average or below-average financial literacy can utilise artificial intelligence and machine learning models to predict prices, thereby enabling them to make risk-averse investments on the stock exchange.

## 3. Proposed Framework

This section outlines the comprehensive framework adopted to forecast stock prices using transformer-based deep learning models within the context of financial literacy enhancement. The framework integrates theoretical foundations with practical implementation steps to address the inherent volatility and complexity of financial time series. It encompasses the selection and customization of transformer architectures, the incorporation of relevant financial indicators, and the application of explainable artificial intelligence (XAI) techniques to ensure transparency in model predictions. The proposed approach is designed not only to improve predictive accuracy but also to support the interpretability of complex machine learning models, thereby aligning with the broader objective of promoting informed decision-making among investors.

### 3.1. Theoretical Background

A substantial body of research has indicated that a significant proportion of the population has limited access to reliable financial education and the tools required for effective personal finance management. This dearth of financial literacy gives rise to a range of issues, including suboptimal savings rates, ineffective budgeting, and a paucity of awareness with regard to investment opportunities. In the long term, these issues give rise to suboptimal financial decisions that can have deleterious consequences [17]. In principle, it is hypothesised that individuals' asset allocation decisions are considerably influenced by their socio-demographic and behavioural characteristics, as well as their level of financial literacy. Furthermore, financial literacy has been demonstrated to play a significant mediating role in investment diversification, with increases in age, education, and income levels having a positive effect on diversification in a number of studies [20].

The correlation between financial literacy and participation in the stock market has been demonstrated to be an inverse one; that is to say, low financial literacy has been shown to act as a barrier to market participation, while high financial literacy has been demonstrated to pave the way for financial independence and to shape saving and investment behaviours. These behaviours are affected by economic cycles, technological developments, and generational differences. Furthermore, individuals with limited financial literacy are more susceptible to financial fraud. Consequently, financial literacy fosters individuals' capacity to make informed investment decisions, thereby providing a safeguard against financial fraud [3].

It is evident that a significant body of research has examined the intermediary effect of financial technology (FinTech) on the relationship between artificial intelligence (AI) and financial decision-making. It is important to note that the scope of this relationship encompasses various domains, including natural language processing (NLP), machine learning algorithms, computer vision, predictive analytics, robotic process automation (RPA), blockchain technology, and deep learning. A substantial body of research has been conducted on this subject, and it has been confirmed that AI applications in FinTech play a crucial mediating role in financial decision-making. This finding is particularly significant for banking professionals, who have provided invaluable insights into their experiences and perspectives [4], [21].

Machine learning has been identified as the most effective AI technique, advancing beyond traditional analytical methods to enable more informed decision-making through sophisticated data analysis and pattern recognition. The impact of artificial intelligence (AI) in the financial sector is steadily increasing, providing significant benefits to financial institutions in terms of efficiency, risk management, and improving the customer experience, particularly for those who may exhibit a lack of financial literacy. The role of FinTech in this regard is to serve as an intermediary between AI and financial decision-making, analysing extensive data sets to provide decision-makers with actionable insights. These technologies have been demonstrated to enhance operational processes, improve risk assessment, and facilitate faster and more accurate decisions [5], [22].

In this context, the study [23] proposed the development of a prototype self-guided stock investment platform that serves as a comprehensive information source for novice investors interested in participating in stock market investments. The objective of this study is to assist users in making informed decisions by providing a dashboard that offers key stock metrics and sector insights to enhance financial literacy, as well as to improve investment decision-making processes by offering predictive modelling using a hybrid LSTM-GARCH model. It is hypothesised that this model will facilitate decision-making processes by combining business intelligence tools with predictive modelling algorithms, thereby increasing the financial literacy of novice investors.

Explainable Artificial Intelligence (XAI) facilitates a more nuanced comprehension of price fluctuations in financial markets, operating under the premise of the Efficient Market Hypothesis (EMH). In particular, the strong and semi-strong form tests of the EMH, which are based on information transparency, can be enhanced by XAI's algorithmic analysis capabilities. The extent to which publicly available information is reflected in prices, and the impact of insider information, can be modelled through this technology's data-driven approach to market dynamics [31].

Furthermore, Explainable Artificial Intelligence (XAI) provides data-driven analyses and forecasts to support investment decisions. To illustrate this point, consider XAI-based systems that function in conjunction with approaches such as the Dow Theory and the Elliott Wave Theory. These systems analyse market trends and wave structures to transparently reveal their effects on investor trends [32]. In particular, the Dow Theory's trend analysis-based processes, such as accumulation, participation, and exhaustion phases, have become testable thanks to XAI's advanced data processing capabilities.

In addition, XAI's capacity for anomaly detection has been shown to facilitate the early identification of bubbles and irrational price movements in financial markets. The utilisation of AI (Artificial Intelligence) models in conjunction with established technical analysis methods, such as Fibonacci ratios, has been posited as a means of mitigating risk by virtue of its ability to facilitate the timely detection of market anomalies [30]. Furthermore, the potential exists for the optimisation of market risk management through the reduction of errors caused by irrational human factors.

In conclusion, XAI should be regarded as a tool that has the capacity to provide decision support mechanisms and to educate users in financial markets. These models facilitate the comprehension of the fundamental mechanisms of algorithmic processes by investors, thereby enabling them to make more informed decisions. For instance, analytical processes that transparently explain the differences between theoretical prices and market prices in futures markets can be demonstrated [6]. In this regard, the innovative solutions offered by XAI and Fin-Tech technologies in increasing financial literacy and supporting informed investment decisions have the potential to bring about significant transformation at the individual and institutional levels, supporting individuals in situations such as risk avoidance and optimal financial decision-making with minimal financial literacy.

### 3.2. Application Framework

The application framework presents the practical implementation of the proposed methodology, detailing the specific components and processes involved in building and evaluating the forecasting models. This includes the selection of appropriate transformer architectures, the systematic collection of historical financial data from the TradingView platform, and the engineering of a comprehensive feature set through the integration of technical indicators. Additionally, this section describes the experimental design, model training procedures, and evaluation metrics employed to assess performance. The application framework is structured to ensure reproducibility and to demonstrate how advanced machine learning techniques can be leveraged to facilitate financial literacy through accessible and interpretable forecasting tools.

#### 3.2.1. Transformer models

**Hybrid deep learning time series forecasting model (LSTNet)** [13] is a hybrid deep learning architecture that captures both short-term local dependencies and long-term periodic patterns in multivariate time series data. It integrates convolutional neural networks, recurrent neural networks, and an autoregressive component into a unified forecasting framework. A convolutional layer is used to extract local and short-term temporal features, identifying high-frequency variations in the input sequence. The extracted features are then passed to a recurrent layer, typically composed of Gated Recurrent Units (GRUs), which models long-term temporal dependencies and evolving patterns. To improve the model's capacity to capture seasonal trends, skip connections provide access to hidden states at fixed intervals. Additionally, an autoregressive component is used to handle residual linear dependencies and abrupt shifts, enhancing robustness in non-stationary environments. Finally, the outputs from both the recurrent and AR components are fused to form the final prediction. This multi-path architecture allows LSTNet to leverage both deep feature learning and interpretable statistical modeling, making it highly suitable for applications with complex temporal patterns.

**Transformer-based time series forecasting model (TST)** [35] is a transformer-based architecture developed for multivariate time series forecasting and representation learning. It adapts the original Transformer model, which is initially designed for natural language processing, to effectively capture temporal dependencies in sequential numerical data. Each time step and variable is linearly embedded into a dense vector space, and sinusoidal positional encodings are added to maintain the temporal order. The model uses multi-head self-attention to capture contextual relationships across time steps and variables, enabling it to detect both short-term interactions and long-range dependencies. To preserve causality in autoregressive forecasting, masking is applied during training to prevent the model from accessing future time steps. Each Transformer block includes residual connections, normalization layers, and feedforward networks to enhance stability and learning capacity. The final output is produced through a projection layer that maps hidden states to the target values, supporting multi-step forecasting tasks. TST's attention-based design offers flexibility and strong generalization capabilities across diverse time series datasets.

**Transformer-based sequence modeling (Vanilla)** [28] provides the foundational architecture for attention-based sequence modeling. While originally intended for machine translation, its ability to model long-range dependencies without recurrence has led to its adoption in time series forecasting. In this framework, each input token is embedded into a continuous vector space, with added positional encodings to retain sequence order. Multi-head self-attention enables each time step to attend to all others in the sequence, capturing complex interdependencies across the input. Position-wise feedforward networks enhance non-linearity and model capacity at each step. Residual connections and layer normalization ensure stable gradient propagation and improved training dynamics. The architecture typically consists of an encoder-decoder structure, where the encoder captures input sequence dependencies and the decoder generates output predictions by attending to both previous outputs and the encoded input. Despite its flexibility and representational power, the quadratic complexity of the attention mechanism can be a limitation for very long time series sequences.

**Decomposition-based linear time series forecasting (DLinear)** [34] is a lightweight and interpretable forecasting model that adopts a decomposition-based linear approach. Unlike deep neural models that rely on large parameter spaces, DLinear assumes that time series behavior can be effectively modeled by applying linear operations to decomposed components such as trend and seasonality. The model first decomposes the input sequence using simple moving averages to isolate different temporal components. Each component is then modeled separately using a dedicated linear layer, capturing component-specific linear patterns. This decomposition strategy allows the model to reduce computational overhead while preserving accuracy. After processing, the individual component predictions are aggregated to reconstruct the overall time series output. DLinear's simplicity enables efficient training and inference, making it well-suited for real-time forecasting scenarios. Its interpretable structure also allows for easier understanding of the contribution of different temporal elements to the final prediction.

### 3.2.2. XAI methods

**SHapley Additive exPlanations (SHAP)** [16] is a unified framework for interpreting machine learning model predictions through consistent and locally accurate feature attribution values. Based on cooperative game theory and the concept of Shapley values, SHAP quantifies the contribution of each feature to the prediction by computing how much each feature adds to the difference between the model's output and its expected output. This approach satisfies several desirable properties including efficiency, symmetry, and additivity. SHAP also unifies a range of existing interpretability methods, demonstrating that popular techniques such as LIME, DeepLIFT, and Layer-Wise Relevance Propagation are specific cases of Shapley value estimation under certain assumptions. The framework is model-agnostic, meaning it can be applied to any predictive model without requiring access to internal parameters or gradients. By ensuring that the sum of feature contributions equals the model's prediction shift, SHAP provides a complete and interpretable decomposition of the model's decision process. Specialized variants such as TreeSHAP, DeepSHAP, and KernelSHAP further extend the framework's applicability to different model types while preserving theoretical consistency.

**Local Interpretable Model-agnostic Explanations (LIME)**, introduced by Ribeiro et al. [25], is a model-agnostic technique for interpreting individual predictions by approximating complex models locally with simpler, interpretable models. It operates on the principle that while the global behavior of a model may be highly non-linear and complex, its behavior in the vicinity of a single instance can often be approximated with a linear or decision tree-based surrogate. To generate local explanations, LIME perturbs the original input to create synthetic neighbors, observes the resulting prediction changes, and fits a simple model to this neighborhood to identify the most influential features. This process results in an interpretable representation that highlights which features most affect the specific prediction. The technique balances explanation fidelity with simplicity by promoting sparse models that are easier for humans to understand. LIME is particularly useful in high-stakes domains where understanding individual predictions is critical, and it provides actionable insights that enhance transparency without requiring access to the internal workings of the predictive model.

### 3.2.3. Data collection

In this study, a dataset is constructed by collecting historical stock market data for five major Turkish banks listed on Borsa Istanbul (BIST): Akbank (AKBNK), QNB Finansbank (QNBF), Garanti Bank (GARAN), İşbank (ISCTR), and Vakıfbank (VAKBN). In addition to individual stock data, two principal Borsa Istanbul indices—XBANK (banking sector index) and XU100 (BIST 100 index)—are incorporated into the dataset to provide a broader market perspective and to evaluate model performance on aggregated financial instruments. The dataset spans a ten-year period from January 2015 to March 2025 and is retrieved via the TradingView API, which grants access to comprehensive daily trading data. The dataset includes key financial indicators commonly used in quantitative equity analysis. These consist of the opening price, the highest and lowest prices recorded during the trading day, the last price at market close, the total trading volume for the day, and the corresponding date and time for each entry. These variables collectively represent market activity and serve as primary inputs for forecasting and modeling tasks. It should be noted that Borsa Istanbul remains closed on weekends and national public holidays. As a result, although a calendar year comprises 365 days, the actual number of trading days is considerably lower. During the dataset construction process, only valid trading days on which the exchange is operational are included. To maintain temporal consistency and ensure data integrity, non-trading days such as weekends and official holidays are systematically excluded. This filtered and chronologically ordered dataset serves as a reliable foundation for developing forecasting models and analyzing long-term financial trends within the Turkish banking sector and broader market indices.

**Table 1**
A comprehensive list of technical indicators utilized in the dataset

| Indicator | Explanation |
| --- | --- |
| EMA_25 | Exponential Moving Average over a 25-day window. Recent prices are weighted more heavily using an exponential function. |
| EMA_50 | Exponential Moving Average over a 50-day window, useful for identifying intermediate trends. |
| EMA_100 | Exponential Moving Average over a 100-day window, often used to track medium-term market movements. |
| EMA_200 | Exponential Moving Average over a 200-day window, commonly used as a long-term trend indicator. |
| EMA_300 | Exponential Moving Average over a 300-day window, representing extended trend movements. |
| RSI_14 | Relative Strength Index calculated over a 14-day window, used to measure the magnitude of recent price changes and identify overbought or oversold conditions. |
| ATR_14 | Average True Range over 14 days, indicating market volatility by averaging the true range over the period. |
| BB_Middle | Middle line of the Bollinger Bands, typically a 20-day simple moving average. |
| BB_Upper | Upper Bollinger Band, located two standard deviations above the middle band. |
| BB_Lower | Lower Bollinger Band, located two standard deviations below the middle band. |
| Tenkan_Sen | Conversion Line of the Ichimoku system; calculated as the average of the highest high and lowest low over the last 9 periods. |
| Kijun_Sen | Base Line of the Ichimoku system; computed as the average of the highest high and lowest low over the last 26 periods. |
| Senkou_Span_A | Leading Span A; average of Tenkan_Sen and Kijun_Sen, plotted 26 periods ahead. |
| Senkou_Span_B | Leading Span B; average of the highest high and lowest low over the past 52 periods, plotted 26 periods ahead. |
| Chikou_Span | Lagging Span; today's closing price plotted 26 periods back. |

In this study, a diverse set of technical indicators is computed in addition to the retrieved market data, using the Python pandas library. These indicators, which are widely employed in financial forecasting, are designed to capture key characteristics of time series data such as trend direction, momentum, and volatility. Each indicator is calculated independently using custom Python scripts based on the fundamental columns: open, high, low, close, and volume. The selection of indicators is grounded in both theoretical and empirical considerations, with an emphasis on their effectiveness in detecting price momentum, trend strength, volatility levels, and potential reversal patterns. Core market variables—including open, close, and volume—obtained via an API, serve as the primary inputs for the computation of these technical indicators. Custom functions written in Python are employed to generate widely used metrics such as the Exponential Moving Average (EMA) and the Relative Strength Index (RSI), which are instrumental in identifying momentum shifts and overbought or oversold market conditions. Additionally, more advanced indicators—such as Bollinger Bands, Average True Range (ATR), and elements of the Ichimoku Kinko Hyo system—are integrated into the dataset to provide insights into volatility patterns, trend behaviors, and reversal signals. These enhanced features allow the modeling framework to capture deeper temporal dependencies and improve the predictive performance of transformer-based architectures. A detailed summary of each technical indicator is presented in Table 1. Their inclusion contributes to a richer temporal representation of stock dynamics, thereby enabling models such as DLinear, LSTNet, TST, and Vanilla to detect more meaningful patterns and enhance forecasting accuracy.

### 3.2.4. Methodology

In this study, a robust forecasting model is developed to predict stock prices using time series data from bank stocks traded on Borsa Istanbul. For this purpose, a multidimensional dataset is constructed by combining historical market data with a comprehensive set of technical indicators. A time series-based transformer approaches are employed to model complex temporal patterns within the data. The overall architecture of the proposed methodology is illustrated in Figure 1.

The input data used in the model consists of daily market variables such as open, high, low, close, and volume, as well as a variety of computed technical indicators. These features are preprocessed using the pandas library and formatted appropriately for time series forecasting. To ensure that the model learns effectively and is not biased by differences in scale across features, all variables are normalized to the [0, 1] range using min-max normalization prior to training. The dataset is divided into training and testing subsets based on predefined split ratios. For each dataset used in this study, a fixed train-test split ratio of 80%–20% is applied to ensure consistent and fair evaluation. To minimize learning bias caused by the sequential nature of the data, the training set is shuffled once before the training process begins. This initial randomization helps the model generalize better by exposing it to varied input sequences during training and reduces the likelihood of overfitting. The sequence length parameter defines the number of previous time steps the model uses as input to make future predictions. It plays a crucial role in balancing the model's ability to capture both short-term fluctuations and longer-term dependencies. A short sequence length might miss essential historical context, whereas overly long sequences can introduce noise and increase model complexity unnecessarily. Therefore, identifying an optimal sequence length is vital for achieving high forecasting accuracy. In this study, multiple values of sequence length are systematically tested for each model, and the most suitable value is selected based on performance metrics such as MSE, RMSE, and MAPE.

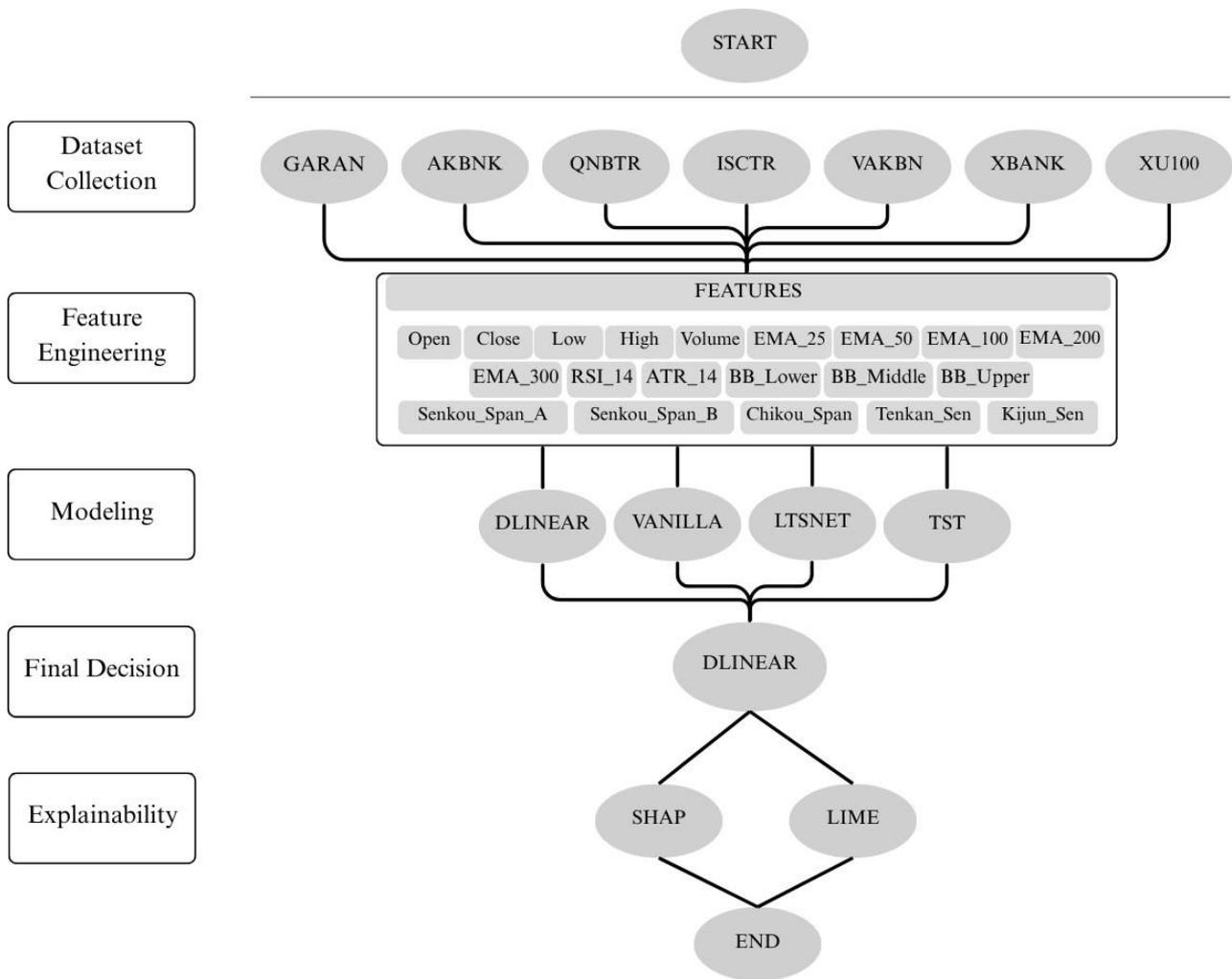

**Figure 1:** Flowchart of the proposed framework

For the DLinear model, an extensive hyperparameter tuning process is carried out to ensure optimal forecasting accuracy and stable training dynamics. Several key parameters—including the number of epochs, batch size, learning rate, sequence length, and dropout rate—are systematically varied and evaluated to identify the most effective configuration. The number of epochs is tested over a range of 50, 100, and 150 to explore the trade-off between training duration and model generalization. It is observed that while longer training (150 epochs) occasionally improves training loss, it also increases the risk of overfitting, particularly in combination with small learning rates. Conversely, shorter training durations (e.g., 50 epochs) lead to underfitting in certain configurations, suggesting insufficient model exposure to temporal patterns. A training duration of 100 epochs is found to provide the best balance between convergence and generalization. Batch size, which affects both training stability and computational efficiency, is varied across 16, 32, and 64. Smaller batch sizes such as 16 are noted to increase gradient variance and result in noisy loss curves, making convergence less reliable. On the other hand, very large batch sizes like 64 tend to produce flatter training curves but sometimes lead to poorer generalization on unseen data. The model demonstrates the most consistent and robust performance with a batch size of 32. Learning rate plays a critical role in shaping the convergence behavior of DLinear. Values of 1e-3, 1e-4, and 1e-5 are explored. A high learning rate (1e-3) facilitates fast convergence, while a very low rate (1e-5) leads to slow and unstable learning, often causing the model to settle into suboptimal local minima or exhibit symptoms of overfitting. Mid-range values (1e-4) offer improved stability, but the overall best results in terms of convergence speed and generalization are achieved with a learning rate of 1e-3. The dropout rate is also varied to examine its regularization effect. Values of 0.0, 0.1, 0.2, and 0.3 are tested. Although higher dropout values (0.2 and 0.3) can mitigate overfitting in deeper architectures, they are found to negatively impact performance in DLinear, which is inherently lightweight. Introducing dropout in this setting often distorts the temporal structure captured by the model, particularly when combined with low learning rates. The best results are consistently obtained when no dropout is applied (dropout = 0.0), reinforcing the model's reliance on its simple linear components for effective learning. Sequence length, which determines the number of previous time steps the model considers for each prediction, is also tested across values of 5, 10, 30, and 60. The model achieves its best performance with a sequence length of 10, which strikes the optimal balance between capturing short-term dependencies and maintaining simplicity in input structure. In summary, the optimal hyperparameter configuration for the DLinear model is determined to be 100 training epochs, a batch size of 32, a learning rate of 1e-3, a sequence length of 10, and a dropout rate of 0. This setup yields the most stable training curves, minimizes error metrics, and avoids overfitting, thereby enabling the model to capture meaningful temporal structures in stock price data with high predictive accuracy.

For the Vanilla Transformer model, hyperparameter tuning is conducted over a broad range of values to assess the sensitivity of the model to training duration, batch size, learning rate, sequence length, and dropout rate. Epoch values of 30, 40, 50, and 60 are tested to determine the optimal duration for convergence. While 30 epochs lead to underfitting and 60 sometimes cause overfitting in volatile datasets, 50 epochs strike the best balance between convergence speed and generalization. Batch sizes of 32, 64, and 128 are examined. A batch size of 32 introduces high variance in gradients, leading to noisy loss behavior, whereas 128 tends to overly smooth the learning dynamics and occasionally impairs generalization. The model performs most reliably at a batch size of 64, where both training stability and validation performance are maximized. Learning rates of 1e-3, 1e-4, and 1e-5 are tested. A learning rate of 1e-3 accelerates convergence but increases the risk of overshooting and model instability. The lowest rate, 1e-5, results in slow learning and early stagnation. The best results are consistently achieved with a learning rate of 1e-4. Dropout rates of 0.1, 0.2, and 0.3 are tested to regularize the model and prevent overfitting. Although higher dropout improves generalization slightly in early epochs, excessive regularization degrades the model's ability to capture sequence-level dependencies. A dropout rate of 0.1 yields the best performance by preserving model expressiveness while still mitigating overfitting. In terms of sequence length, values of 5, 10, 30, and 60 are evaluated. The model achieves the best performance with a sequence length of 10, effectively capturing relevant temporal dynamics without introducing unnecessary complexity. Overall, the optimal configuration for the Vanilla Transformer model consists of 50 training epochs, a batch size of 64, a learning rate of 1e-4, a sequence length of 10, and a dropout rate of 0.1.

For the TST model, a Transformer-based architecture specialized for time series, hyperparameters are tuned systematically to optimize predictive accuracy. The number of epochs is varied across 30, 40, 50, and 60. A training duration of 50 epochs is found to offer a good trade-off between learning depth and resistance to overfitting, especially in combination with appropriate regularization. Batch sizes of 16, 32, and 64 are compared. The model shows instability and variance spikes with a batch size of 16, while 64 occasionally leads to slower convergence. A batch size of 32 offers the most balanced and reliable training performance. Learning rate experiments involve values of 1e-3, 1e-4, and 1e-5. As with other Transformer-based models, 1e-3 converges rapidly but can destabilize training. 1e-5 results in sluggish learning. The best outcomes are observed at 1e-4, which maintains stable gradients and strong generalization. The model's dropout rate is also adjusted between 0.0 and 0.3. Dropout rates of 0.2 and 0.3 are found to hamper sequence encoding fidelity, while a 0.0 dropout setting increases the risk of overfitting. A dropout value of 0.1 provides the best compromise, enabling generalization without degrading sequence representation. Regarding the sequence length, the model is tested on 5, 10, 30, and 60 time steps. The best predictive performance is observed with a sequence length of 5, suggesting that the model benefits from more recent temporal information when identifying patterns. Hence, the best configuration for the TST model consists of 50 epochs, a batch size of 32, a learning rate of 1e-4, a sequence length of 5, and a dropout rate of 0.1.

In the case of LSTNet, a hybrid model incorporating convolutional, recurrent, and autoregressive components, hyperparameter optimization is particularly important due to the model's architectural complexity. Epoch values of 50, 80, 100, and 150 are tested. A training duration of 100 epochs is found to yield the most stable and accurate predictions, with 150 epochs increasing the likelihood of overfitting and 50 often leading to undertrained models. Batch sizes of 32, 64, and 128 are evaluated. Smaller sizes increase computational time and noise, while the largest size occasionally causes underfitting. The best trade-off between model convergence speed and generalization is observed at a batch size of 64. Learning rate tuning across 1e-4, 1e-5, and 1e-6 reveals that lower rates improve stability and reduce overfitting. Although 1e-6 sometimes slows convergence too much, 1e-5 provides the most reliable overall performance. Dropout rates from 0.0 to 0.3 are tested to regulate overfitting in the recurrent and CNN layers. A dropout rate of 0.2 effectively balances regularization and model expressiveness, particularly in long sequence inputs where temporal abstraction is

**Table 2**

Experimentally optimized parameter configurations for each transformer model

| Models  | Epochs | Learning Rate | Batch Size | Sequence Length | Dropout |
|---------|--------|---------------|------------|-----------------|---------|
| DLinear | 100    | 1e-3          | 32         | 10              | 0       |
| Vanilla | 50     | 1e-4          | 64         | 10              | 0.1     |
| TST     | 50     | 1e-4          | 32         | 5               | 0.1     |
| LTSNet  | 100    | 1e-5          | 64         | 5               | 0.2     |

**Table 3**

Evaluation of DLinear model performance for each stock

| Stock | MSE        | MAE      | MAPE (%) | RMSE     | $R^2$  |
|-------|------------|----------|----------|----------|--------|
| AKBNK | 2.0925     | 1.0292   | 2.3571   | 1.4465   | 0.9932 |
| GARAN | 6.1836     | 1.7642   | 2.3362   | 2.4867   | 0.9955 |
| ISCTR | 0.1046     | 0.2354   | 2.2352   | 0.3234   | 0.9920 |
| QNBTR | 176.8970   | 8.6004   | 3.4871   | 13.3003  | 0.9847 |
| VAKBN | 0.3110     | 0.4155   | 2.3523   | 0.5576   | 0.9897 |
| XBANK | 86546.8203 | 210.6340 | 2.0433   | 294.1884 | 0.9939 |
| XU100 | 23899.1621 | 117.2031 | 1.4201   | 154.5935 | 0.9923 |

**Table 4**

Evaluation of LSTNet model performance for each stock

| Stock | MSE         | MAE      | MAPE (%) | RMSE     | $R^2$  |
|-------|-------------|----------|----------|----------|--------|
| AKBNK | 6.1311      | 1.8154   | 3.9633   | 2.4761   | 0.9800 |
| GARAN | 22.5975     | 3.3899   | 4.3025   | 4.7537   | 0.9836 |
| ISCTR | 0.4204      | 0.4735   | 4.5308   | 0.6484   | 0.9680 |
| QNBTR | 624.9850    | 14.9640  | 5.8021   | 24.9997  | 0.9470 |
| VAKBN | 0.8710      | 0.7093   | 4.0626   | 0.9333   | 0.9716 |
| XBANK | 319088.093  | 401.7886 | 3.8377   | 564.8788 | 0.9780 |
| XU100 | 73595.0625  | 211.4709 | 2.5952   | 271.2841 | 0.9770 |

critical. The sequence length is also explored using values of 5, 10, 30, and 60. LSTNet achieves its best results with a sequence length of 5, suggesting that its hybrid architecture can effectively extract meaningful features from relatively short temporal windows. Thus, the optimal parameter set for LSTNet is determined as 100 epochs, a batch size of 64, a learning rate of 1e-5, a sequence length of 5, and a dropout rate of 0.2.

Across all models, overfitting is addressed through careful adjustment of the dropout rate and other regularization-oriented hyperparameters. Dropout is strategically used to prevent the models from memorizing training data and to encourage generalization. Each model's optimal configuration includes a dropout value that is specifically tuned to maintain expressiveness while reducing overfitting, thus contributing to the development of robust forecasting models. In conclusion, through the careful tuning of essential parameters such as learning rate, batch size, number of epochs, and sequence length, the proposed model achieves stable and highly accurate forecasts for bank stock prices in Borsa Istanbul. As a result of extensive experimentation, the most successful parameter combinations are presented in Table 2.

## 4. Experiment results

To assess the performance of the proposed time series forecasting model, a set of regression-based evaluation metrics is employed. These metrics include Mean Absolute Percentage Error (MAPE), Mean Absolute Error (MAE), Mean Squared Error (MSE), Root Mean Squared Error (RMSE), and the Coefficient of Determination ($R^2$). Each of these metrics provides a unique perspective on the accuracy, magnitude of error, and explanatory power of the model.

The performance evaluation of the DLinear model across different stocks demonstrates remarkably consistent and high-quality predictive capabilities. The model achieves exceptional $R^2$ values ranging from 0.984 to 0.995, indicating that the DLinear architecture can explain between 98.4% and 99.5% of the variance in stock price movements across all tested securities. This high level of explanatory power suggests that the decomposition-based linear approach effectively captures the underlying temporal patterns in financial time series data. Among the individual stocks, GARAN exhibits the strongest overall performance with the highest $R^2$ value of 0.995 and a low MAPE of 2.336%, demonstrating the model's ability to accurately predict this particular stock's price movements. Similarly, ISCTR shows exceptional performance in terms of absolute error metrics, achieving the lowest MSE (0.104) and RMSE (0.323) values, which can be attributed to its relatively stable price range compared to other securities. The XU100 index demonstrates the most precise percentage-based predictions with the lowest MAPE of 1.420%, suggesting that the model performs particularly well on diversified market indices. The variation in MSE and RMSE values across different stocks primarily reflects the inherent differences in price levels and volatility characteristics rather than model performance inadequacies. For instance, XBANK and XU100 show higher absolute error values due to their higher price ranges, but their $R^2$ and MAPE metrics remain competitive. QNBTR presents the most challenging forecasting scenario with the highest MAPE of 3.487% and lowest $R^2$ of 0.984%, yet these values still represent strong predictive performance. Overall, the consistent achievement of MAPE values below 4% across all securities indicates that the DLinear model provides reliable and practically applicable forecasting accuracy for diverse stock price prediction tasks, as summarized in Table 3.

The performance evaluation of the LSTNet model across a range of financial instruments listed on Borsa Istanbul reveals its capacity to deliver robust and consistent forecasting results. The model achieves $R^2$ values ranging from 0.9470 to 0.993, indicating that it explains between 94.7% and 99.3% of the variance in stock price movements across the evaluated assets. This high explanatory power suggests that LSTNet effectively captures both short-term fluctuations and longer-term dependencies in financial time series, in line with its hybrid architectural design. Among the tested stocks, AKBNK and ISCTR demonstrate the most favorable overall performance in terms of both goodness-of-fit and error-based metrics. ISCTR achieves the lowest MSE (0.104) and RMSE (0.323) values, indicating strong accuracy in absolute terms. Similarly, AKBNK reaches a high $R^2$ score of 0.993 with a relatively low MAPE of 2.357%, highlighting its suitability for stable and predictable time series. GARAN also exhibits strong predictive performance with the highest R² of 0.995 and a MAPE of 2.336%, confirming the model's capability to generalize well for this particular security. On the other hand, QNBTR emerges as the most challenging series, characterized by elevated volatility, reflected in the highest MSE (624.985) and RMSE (24.9997), as well as the lowest R² (0.9470). Despite these larger absolute errors, the MAPE remains within an acceptable range (5.8021%), indicating the model's robustness under more volatile conditions. Notably, the model performs well on market indices such as XBANK and XU100. While these indices present higher absolute error values (e.g., RMSE of 564.8788 for XBANK), their MAPE values remain competitively low (2.043% and 2.5952%, respectively), underscoring the model's ability to provide stable percentage-based forecasts even for high-magnitude targets. Overall, the consistent MAPE values below 6% across all securities demonstrate that LSTNet offers dependable forecasting accuracy. Its integrated use of convolutional, recurrent, and autoregressive components proves effective in handling diverse temporal characteristics across various financial assets, as detailed in Table 4.

**Table 5**
Evaluation of Vanilla model performance for each stock

| Stock | MSE | MAE | MAPE (%) | RMSE | R² |
|---|---|---|---|---|---|
| AKBNK | 9.2647 | 2.1298 | 4.7748 | 3.0438 | 0.9702 |
| GARAN | 25.4778 | 3.6587 | 4.7629 | 5.0476 | 0.9814 |
| ISCTR | 0.5773 | 0.5756 | 5.1121 | 0.7598 | 0.9556 |
| QNBTR | 662.3851 | 15.7425 | 6.0481 | 25.7368 | 0.9426 |
| VAKBN | 0.7659 | 0.6841 | 3.9776 | 0.8752 | 0.9747 |
| XBANK | 340070.6566 | 432.0704 | 4.2421 | 583.1558 | 0.9762 |
| XU100 | 104267.3820 | 255.0448 | 3.0523 | 322.9046 | 0.9665 |

**Table 6**
Evaluation of TST model performance for each stock

| Stock | MSE | MAE | MAPE (%) | RMSE | R² |
|---|---|---|---|---|---|
| AKBNK | 9.3008 | 2.1219 | 4.8905 | 3.0497 | 0.9701 |
| GARAN | 28.7642 | 3.6729 | 4.5917 | 5.3632 | 0.9792 |
| ISCTR | 0.4144 | 0.4886 | 4.6287 | 0.6438 | 0.9684 |
| QNBTR | 534.6752 | 15.6350 | 6.9505 | 23.1230 | 0.9547 |
| VAKBN | 1.0669 | 0.7744 | 4.2575 | 1.0329 | 0.9652 |
| XBANK | 345788.6250 | 432.4882 | 4.3094 | 588.0380 | 0.9761 |
| XU100 | 91758.4609 | 227.3230 | 2.8685 | 302.9166 | 0.9713 |

The performance analysis of the Vanilla Transformer model across various financial instruments reveals a competent yet slightly less refined predictive capacity compared to more specialized architectures. The model achieves R² values ranging from 0.9426 to 0.9814, indicating its ability to explain between 94.26% and 98.14% of the variance in stock prices. Although these values reflect strong performance overall, they remain somewhat lower than those achieved by DLinear and LSTNet models, suggesting a relative limitation in capturing temporal intricacies without specialized architectural adaptations. Among the individual stocks, GARAN achieves the highest R² value of 0.9814 with a MAPE of 4.7629%, demonstrating the model's strength in modeling moderately volatile series. VAKBN also yields strong results with low MSE (0.7659) and RMSE (0.8752), along with a relatively low MAPE of 3.9776%, indicating effective prediction in stable market conditions. However, the model shows reduced precision for highly volatile stocks. For example, QNBTR exhibits the highest MAPE (6.0481%) and the lowest R² (0.9426), accompanied by a high RMSE of 25.7368. These metrics suggest that the Vanilla Transformer struggles to generalize effectively for stocks with wider fluctuations unless further optimized. Similarly, ISCTR, despite its low absolute error metrics, records a relatively high MAPE of 5.1121%, reflecting sensitivity to small price movements. On aggregate indices such as XBANK and XU100, the model handles percentage-based predictions well despite elevated absolute errors due to high nominal values. The XU100 index, for instance, reaches a MAPE of only 3.0523% even though its RMSE is over 322. These outcomes indicate the model's robustness for broader market tracking, especially when normalization or scaling is properly applied. In summary, while the Vanilla Transformer delivers reliable and interpretable results across all tested securities, its performance is slightly constrained by the lack of domain-specific enhancements. The model remains a solid baseline for comparison and a valuable foundation for further architecture-specific improvements, as outlined in Table 5.

The evaluation of the Time Series Transformer (TST) model across multiple Borsa Istanbul-listed financial instruments highlights its strong and consistent predictive capability. The model yields R² values ranging from 0.9547 to 0.9792, reflecting its ability to explain over 95% of the variance in price movements for all tested assets. These results indicate that TST effectively captures both temporal dependencies and cross-feature interactions, even in the presence of financial time series volatility. Among the individual assets, ISCTR achieves the best absolute accuracy with the lowest MSE (0.4144) and RMSE (0.6438), confirming the model's precision in relatively stable market conditions. Meanwhile, the XU100 index displays the lowest MAPE (2.8685%) despite high absolute error values, underscoring the model's robustness in capturing percentage-based accuracy across diversified indices. Stocks such as GARAN and AKBNK also demonstrate solid performance, each maintaining R² scores above 0.97 with MAPE values under 5%, validating the model's adaptability to a variety of market behaviors. In contrast, QNBTR poses a more difficult challenge, yielding the highest MAPE (6.9505%) and RMSE (23.1230), which can be attributed to its pronounced volatility. However, even in this case, the R² value remains relatively high at 0.9547, indicating that the model still captures the general directional movement effectively. Performance on index-based series such as XBANK and XU100 remains competitive. Despite their large price scales, TST maintains low MAPE values (4.3094% and 2.8685%, respectively), reflecting its scalability and effectiveness on high-value assets. In conclusion, the TST model delivers dependable forecasting results across a diverse set of securities. Its performance, as summarized in Table 6, demonstrates the effectiveness of attention-based architectures in time series applications, particularly in capturing both short-term fluctuations and long-range dependencies.

A comprehensive comparison of forecasting performance across all evaluated models—including DLinear, LSTNet, Vanilla Transformer, and TST—reveals that the DLinear architecture consistently delivers the most accurate and reliable predictions. It achieves the highest $R^2$ values (up to 0.995) and the lowest mean absolute and percentage errors across most stocks, indicating superior generalization and minimal deviation from actual market behavior. Notably, DLinear also maintains remarkably stable performance across both volatile and stable stocks, demonstrating its robustness and scalability in handling diverse financial time series patterns. These findings underline the effectiveness of decomposition-based linear modeling in isolating trend and seasonal components, allowing DLinear to outperform more complex deep learning and attention-based models in terms of both absolute and relative error metrics. Given its high explanatory power and computational efficiency, DLinear emerges as the most practical and interpretable choice for real-world financial forecasting applications.

Figures 2 and 3 provide two complementary visualizations highlighting the training dynamics and prediction performance of the DLinear model. During the experiments, loss curves for all XBANK and XU100 indices, along with five stocks, exhibited highly similar models characterized by smooth and stable trajectories that decreased with minimal variation on a period-by-period basis. To avoid repetition and enhance clarity, Figure 3 presents only the loss curve corresponding to the stock with the best overall performance (GARAN). This representative example effectively demonstrates the model's convergence behavior and provides sufficient information about learning stability in a broader dataset. In contrast, Figure 2 displays the predicted versus actual stock prices for all stocks. These plots visually demonstrate the model's forecasting capability across a diverse set of securities. The close alignment between predicted and observed values in each case confirms the DLinear model's ability to capture underlying trends and adapt to varying market conditions. Together, these figures offer a comprehensive perspective on both the optimization process and the generalization strength of the DLinear model in multi-stock forecasting tasks.

Figure 3 illustrates the epoch-wise training and testing loss curves of the DLinear model applied to the GARAN stock, which yielded the best overall performance among all evaluated securities. The plot demonstrates a rapid initial decrease in both training and testing loss within the first few epochs, indicating that the model quickly learns the underlying data distribution. After approximately the 10th epoch, both curves converge and stabilize near zero, suggesting that the model reaches an optimal solution early in the training process. The close alignment between the training and testing loss curves further confirms that the model generalizes well and avoids overfitting. The absence of any divergence between the two curves implies consistent learning across both seen and unseen data. Moreover, the flat and minimal loss values throughout the remainder of the training phase underscore the model's robustness and stability when applied to the GARAN time series. To complement the observed predictive stability, it is crucial to investigate not only how the model generates accurate forecasts but also why it arrives at specific predictions. In the context of financial time series, where transparency and interpretability are of paramount importance, understanding the internal decision logic of the forecasting model becomes essential. To this end, explainable AI (XAI) techniques—specifically SHAP and LIME—are employed to elucidate the contribution of individual features to the model's output. The subsequent section presents global and local interpretability analyses for selected stocks, offering deeper insights into the model's feature attribution patterns and decision-making behavior. Figure 4 presents the explainability analysis for the AKBNK stock forecasting model. The SHAP summary plot, which provides a global perspective by averaging feature contributions across the dataset, indicates that the model heavily relies on momentum indicators, particularly the `RSI_14` values at time steps `t-4`, `t-8`, and `t-7`. These are followed by features such as `MACD_Signal` and `Volume`, highlighting a consistent dependency on short- and mid-term price dynamics throughout the time series. In contrast, the LIME local explanation—applied to the final data instance—reveals a different pattern of feature importance. In this specific case, the prediction is primarily influenced by volatility and trend-based indicators, including `Kijun_Sen_t7`, `ATR_14_t5`, `EMA_50_t6`, and `BB_Middle_t9`. Notably, while `MACD_Signal_t5` contributes positively in both explanations, several features with high LIME values—such as `Senkou_Span_A`, `Chikou_Span`, and `Tenkan_Sen`—do not appear prominently in the SHAP ranking. This divergence between global and local interpretations underscores the model's context-sensitive behavior: while momentum-based features dominate on average, the model adapts by shifting its attention to longer-horizon trend and volatility signals in response to specific market conditions. It is important to note that all LIME-based local explanations presented in this study correspond to the same data point: the final day available in each respective dataset. Since LIME operates on a single instance at a time, the selected instance for interpretability across all assets and indices is consistently the most recent observation in the time series. This ensures uniformity in local interpretability and allows for direct comparison between LIME results across different financial instruments under identical temporal conditions.

Figure 5 illustrates the interpretability assessment for the GARAN stock prediction model. The SHAP feature importance plot provides a global overview by averaging the contribution of each feature across all instances. Here, the model's predictions are primarily influenced by `RSI_14`, particularly at time steps `t-8` and `t-9`, followed by several `Volume` features. This indicates that, at a global scale, the model predominantly relies on momentum-based and volume-related signals to estimate future price movements. On the other hand, the LIME explanation—derived from the final data instance—reveals a different set of influential features. In this local context, the model's decision is largely driven by long-term trend indicators such as `EMA_200_t8`, `EMA_100_t8`, and `BB_Middle_t7`, as well as Ichimoku system components including `Chikou_Span`, `Kijun_Sen`, and `Senkou_Span_A`. These features, while impactful in the specific instance, are largely absent from the SHAP-derived global importance ranking. This divergence underscores the complementarity of global and local explanations: whereas SHAP highlights average dependencies on momentum and volume across time, the LIME analysis indicates that the model adaptively shifts its focus toward trend-following and volatility-sensitive indicators under certain market conditions.

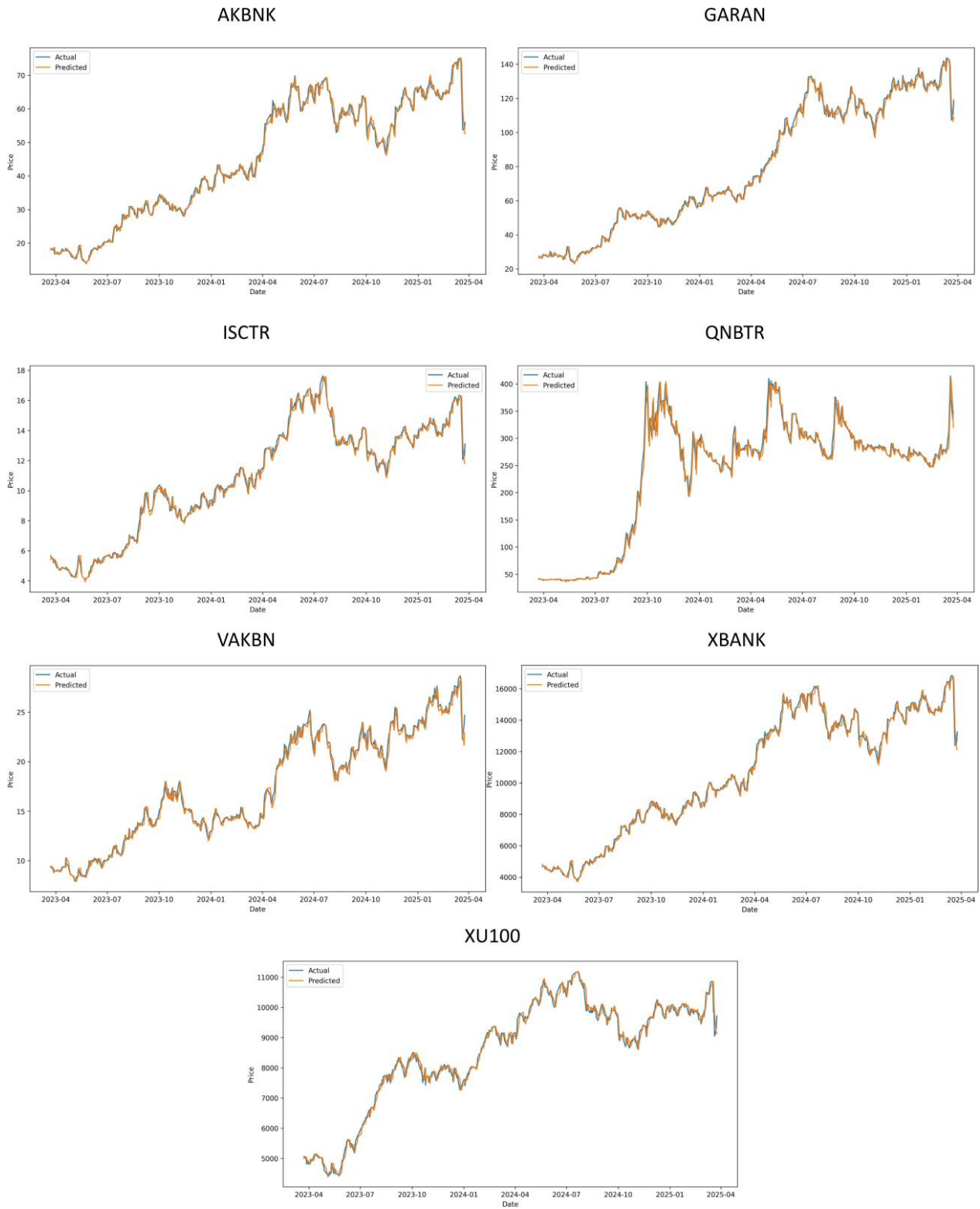

**Figure 2:** Actual and predicted stock values generated by the DLinear model

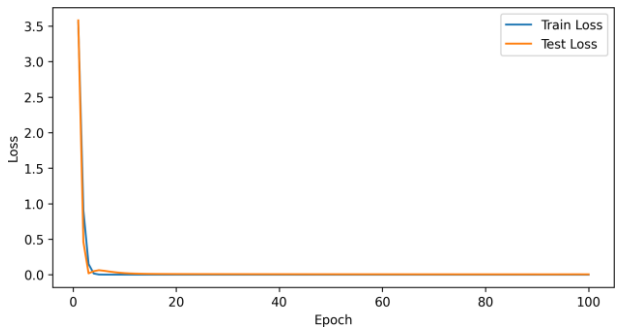

**Figure 3:** Loss trajectory of the DLinear model for GARAN across epoch size

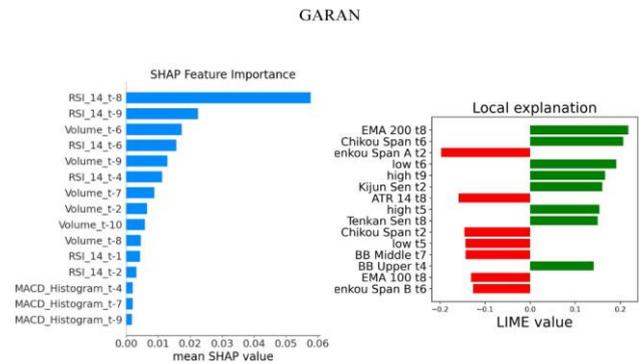

**Figure 5:** Explainable AI analysis of DLinear model predictions for GARAN stock using SHAP and LIME

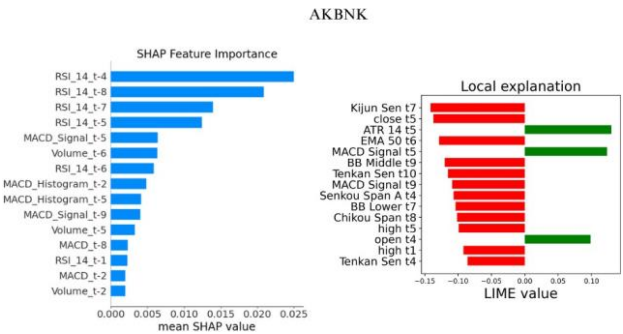

**Figure 4:** Explainable AI analysis of DLinear model predictions for AKBNK stock using SHAP and LIME

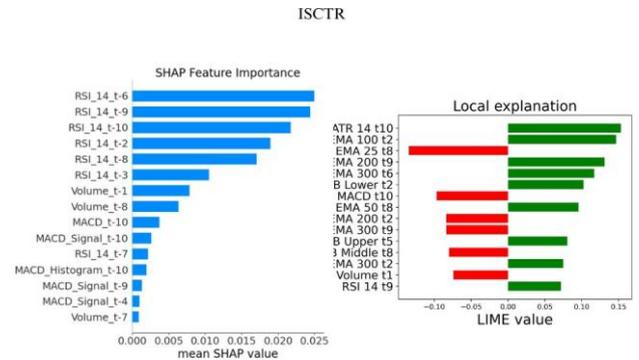

**Figure 6:** Explainable AI analysis of DLinear model predictions for ISCTR stock using SHAP and LIME

Figure 6 shows the interpretability results for the ISCTR stock forecasting model. The SHAP global explanation indicates that the model consistently relies on momentum-based features, with `RSI_14` at multiple time steps (`t-6`, `t-9`, `t-10`, and `t-2`) ranking among the most influential predictors. Volume-related features (e.g., `Volume_t-1`, `Volume_t-8`) and MACD-based indicators also appear in the top contributions, reinforcing the model's preference for short-term price momentum and trading activity when generalized over the dataset. In contrast, the LIME local explanation—focused on the final instance in the test sequence—reveals a dominant influence from trend and volatility-sensitive indicators. Notably, indicators such as `ATR_14_t10`, `EMA_100_t2`, `EMA_25_t8`, and multiple moving averages (e.g., `MA_200`, `MA_300`) appear as key drivers of the prediction. Furthermore, several Bollinger Band components and Ichimoku-derived metrics (e.g., `B_Lower_t2`, `B_Upper_t5`) contribute substantially in this local context, even though they do not appear prominently in the SHAP global summary. This discrepancy highlights the model's context-aware nature: while it relies on momentum indicators on average, it adapts its predictive strategy by incorporating longer-horizon trend and volatility signals when specific structural patterns emerge in the market.

Figure 7 displays the explainability analysis for the QNBTR stock forecasting task. The SHAP plot highlights a dominant role of `RSI_14_t2` and `Volume_t10` in the model's overall predictions, followed by other recent momentum-based inputs such as `RSI_14_t4`, `Volume_t6`, and `RSI_14_t3`. This suggests that, on average, the model primarily depends on short-term RSI values and trading volume across different time lags. On the other hand, the LIME explanation for the final prediction instance presents a different feature hierarchy. In this local context, indicators such as `BB_Middle_t7`, `EMA_200_t10`, and `MACD_Histogram_t2` emerge as major contributors. Additionally, Ichimoku-derived signals like `Kijun_Sen_t7`, `Senkou_Span_B_t9`, and `Chikou_Span_t8` gain significance, despite not being prominent in the SHAP ranking.

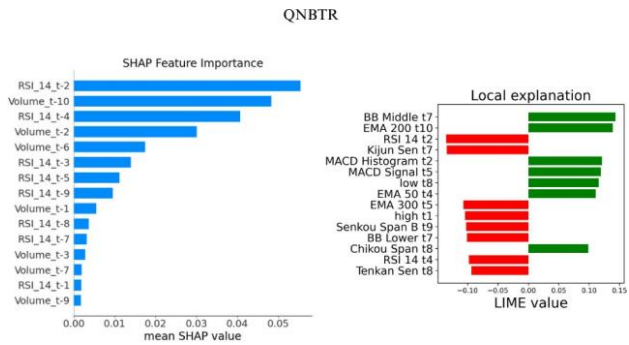

**Figure 7:** Explainable AI analysis of DLinear model predictions for QNBTR stock using SHAP and LIME

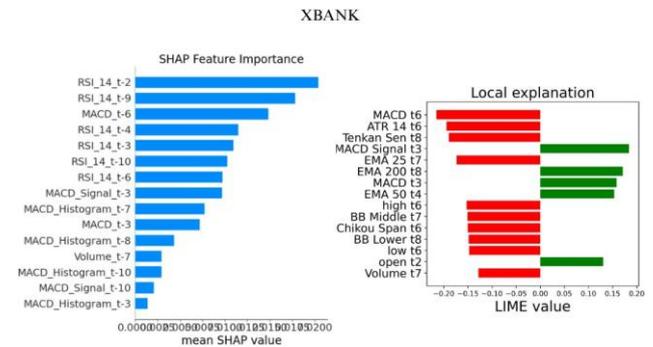

**Figure 9:** Explainable AI analysis of DLinear model predictions for XBANK stock using SHAP and LIME

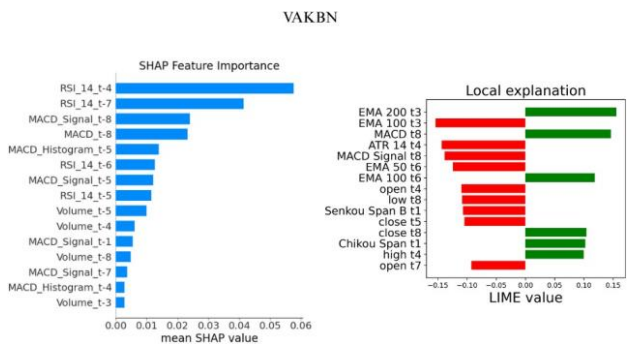

**Figure 8:** Explainable AI analysis of DLinear model predictions for VAKBN stock using SHAP and LIME

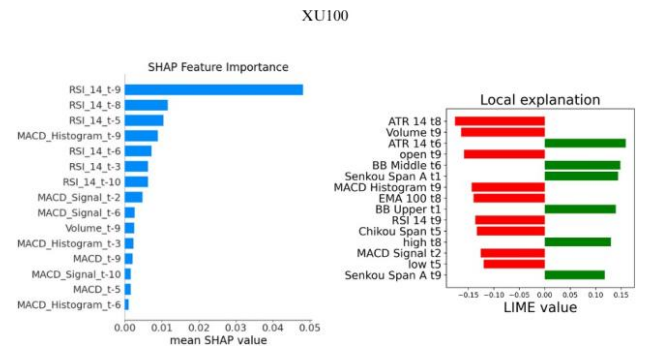

**Figure 10:** Explainable AI analysis of DLinear model predictions for XU100 stock using SHAP and LIME

Figure 8 provides the explainability results for the VAKBN stock forecasting model. According to the SHAP summary plot, the model primarily depends on momentum-related features such as `RSI_14_t4` and `RSI_14_t7`, followed by MACD-based indicators, including `MACD_Signal_t8`, `MACD_t8`, and `MACD_Histogram_t5`. This highlights a strong emphasis on short-term trend and momentum in the model's overall behavior. In contrast, the LIME local explanation highlights the role of longer-term moving averages such as `EMA_200_t3`, `EMA_100_t3`, and `EMA_50_t6`, as well as volatility and price-based inputs like `ATR_14_t4`, `open_t4`, and `close_t8`. Notably, several of these LIME-dominant features—such as `Senkou_Span_B_t1` and `Chikou_Span_t1`—do not appear in the global SHAP importance ranking.

Figure 9 presents the interpretability assessment for the XBANK index forecasting model. The SHAP plot reveals that the model's global behavior is primarily governed by momentum-based indicators, particularly `RSI_14` across various time steps (e.g., `t-2`, `t-9`, `t-6`, and `t-4`). Additionally, MACD-related features, such as `MACD_t6`, `MACD_Signal_t3`, and `MACD_Histogram_t7`, also contribute significantly, suggesting the model consistently exploits trend and momentum signals in its average predictions. In contrast, the LIME explanation for the selected prediction instance prioritizes a different set of features. Here, variables like `EMA_25_t7`, `EMA_200_t8`, and `EMA_50_t4` are among the most impactful, accompanied by volatility-related indicators such as `ATR_14_t6` and Bollinger Band components. While some overlap exists—e.g., `MACD_t3` appears in both explanations—the presence of Ichimoku elements like `Tenkan_Sen_t8` and `Chikou_Span_t6` in LIME but not in SHAP highlights local context dependency in the model's predictive strategy.

Figure 10 presents the interpretability results for the XU100 index forecasting model. The SHAP summary plot indicates a strong global dependence on momentum indicators, especially `RSI_14_t9`, which exhibits the highest mean SHAP value by a wide margin. Other RSI variants at different time steps (e.g., `t-8`, `t-5`, `t-6`) and MACD-related features (e.g., `MACD_Histogram_t9`, `MACD_Signal_t2`) also contribute, though to a lesser extent. In the LIME local explanation, the feature importance distribution shifts toward volatility and trend-based indicators. Features such as `ATR_14_t6`, `BB_Middle_t6`, `EMA_100_t8`, and Ichimoku-related components like `Senkou_Span_A_t1` and `Chikou_Span_t5` appear prominently. Notably, some RSI and MACD inputs, which dominate the global ranking, retain a presence locally but are less influential compared to the broader set of volatility-sensitive signals in this specific instance.

## 5. Discussion

This study investigates the effectiveness of transformer-based deep learning models in forecasting stock prices within an emerging market setting. By focusing on the five highest-volume banking stocks in the BIST100 index, along with the XBANK and XU100 indices, the research provides a comprehensive evaluation of advanced time series forecasting techniques. Among the four transformer models tested—DLinear, LTSNet, Vanilla Transformer, and Time Series Transformer (TST)—DLinear consistently outperforms its counterparts across all performance metrics.

As shown in Table 3, DLinear achieves remarkably low error values (MSE, MAE, RMSE) and high $R^2$ scores, particularly in the case of ISCTR, where the $R^2$ value reaches 0.9920 with a minimal RMSE of 0.3234. Even in more volatile instruments such as XU100, the model demonstrates a robust capacity to generalize patterns, achieving an $R^2$ of 0.9923. These findings underscore the model's adaptability and predictive strength in financial environments marked by high volatility and non-linearity.

Another significant contribution of the study is the application of explainable AI (XAI) methods, namely SHAP and LIME. These techniques not only enhance transparency in model predictions but also facilitate deeper insights into the role of technical indicators in forecasting. This aligns with ongoing efforts to integrate machine learning into financial decision-making processes in a manner that is both interpretable and actionable for investors.

Importantly, the results also highlight the potential of these models to support financial literacy. By demonstrating how complex AI-based forecasting tools can be made accessible and interpretable, the study illustrates a practical pathway for bridging advanced technology and informed investor behavior. This is particularly relevant in emerging markets, where financial education and digital transformation often progress simultaneously. In particular, within the context of Turkey, during periods characterised by the uncertainties surrounding the economic ramifications of seismic events and the implementation of economic policies influenced by electoral considerations, the employment of XAI outputs has the potential to empower investors who are financially illiterate and who lack a theoretical comprehension of price formation. This enables them to manage their processes of selecting the most reliable stocks or reassessing risk aversion criteria and adjusting their positions accordingly in a more optimal manner.

## 6. Conclusion

This study provides compelling evidence on the validity of transformer-based models, particularly DLinear, in forecasting stock prices in an emerging market such as Turkey. Over the past two years, during which the test data were collected, the Turkish economy has undergone a highly dynamic period characterised by high inflation, rising interest rates, and fluctuations in the exchange rate. The conditions described above, compounded by the uncertainties stemming from earthquake and election-driven economic policies, have had an adverse effect on rational decision-making processes. Volatility in energy prices, evolving credit opportunities, and challenges in external financing have also exerted considerable influence on Turkish financial markets. In this volatile macroeconomic environment, Borsa Istanbul has maintained its strategic importance, with key indices such as the high-volume banking sector index (XBANK) and the broader XU100 index serving as critical tools for investor decision-making. It is evident that both financially literate investors and professionals in the finance sector are encountering limitations in their ability to predict future outcomes and position themselves accordingly, as traditional methods become increasingly inadequate. At this juncture, individuals lacking financial literacy are particularly vulnerable, as their capacity to mitigate financial risks is significantly curtailed. The present study demonstrates that, even under such adverse conditions, XAI models provide a framework for reintegrating predictability into financial decision-making processes through cause-effect analyses. The integration of enriched feature sets derived from various technical indicators has been demonstrated to significantly enhance the predictive performance of these models. The combination of transformer models and Explainable Artificial Intelligence (XAI) presents a repeatable and transparent framework for future applications in financial technology and education. The study's methodological rigor and practical focus have yielded several valuable contributions, including:

- The utilisation of sophisticated transformer models in practical financial forecasting scenarios is a subject that merits consideration.
- The enhancement of model interpretability via explainable artificial intelligence is a subject that has attracted much attention in recent years.
- The promotion of financial literacy is to be achieved by rendering complex predictive systems more intelligible and applicable.
- The empirical demonstration of enhanced predictive performance through the augmentation of features has been demonstrated.
- The development of a replicable framework that integrates artificial intelligence with the objectives of financial literacy is of significant importance.

In conclusion, within the context of the complexity of emerging market dynamics, where sudden macroeconomic changes necessitate detailed analysis by investors, XAI contributes significantly. By elucidating market characteristics and rendering intricate data patterns more intelligible, XAI augments the knowledge and analytical capacity of investors. Furthermore, the transparency of XAI has the potential to encourage broader financial participation and to deepen capital markets, particularly in Borsa Istanbul. In volatile economic environments, such as that of Turkey, where there is a high degree of fluctuation in interest rates and exchange rates, XAI-supported models, such as DLinear, reinforced with SHAP and LIME mechanisms, establish a foundation for improved market predictability. The integration of XAI models into the equation has been demonstrated to engender significant advancements in enhancing financial literacy, confidence in investment decisions, and market analysis, thereby paving the way for optimal outcomes.

### 6.1. Research limitations

While the study presents strong empirical results, certain limitations must be acknowledged. First, the dataset is limited to Turkish financial markets, which may affect the generalizability of the findings to other geographical or regulatory contexts. Second, only historical technical indicators are used; macroeconomic variables and sentiment data are not incorporated, which may affect long-term prediction robustness. Third, although explainable AI techniques are applied, their interpretations remain dependent on the quality and relevance of the feature engineering process.

### 6.2. Potential future research

Future studies may explore the integration of macroeconomic indicators, sentiment analysis from financial news or social media, and cross-market transfer learning techniques. In addition, further research could evaluate these models' performance in high-frequency trading environments or during periods of economic crisis. Comparative studies between emerging and developed markets could also provide deeper insights into model adaptability and investor behavior across different financial ecosystems.

## 7. Declaration of Generative AI and AI-Assisted Technologies in the Writing Process

During the preparation of this work the authors used ChatGPT tool in order to improve language and readability. After using this tool, the authors reviewed and edited the content as needed and takes full responsibility for the content of the publication.

## References

[1] Armagan, I.U., 2023. Price prediction of the borsa istanbul banks index with traditional methods and artificial neural networks. Borsa Istanbul Review 23, S30–S39.

[2] Arslankaya, S., Toprak, Ş., 2021. Makine öğrenmesi ve derin öğrenme algoritmalarını kullanarak hisse senedi fiyat tahmini. International Journal of Engineering Research and Development 13, 178–192.

[3] Baveja, G.S., Verma, A., 2024. Impact of financial literacy on investment decisions and stock market participation using extreme learning machines. URL: https://arxiv.org/abs/2407.03498, arXiv:2407.03498.

[4] de Carvalho, M.P.P., 2023. The Impact of Artificial Intelligence on the Banking and Financial Sector's Strategic Decision-Making. Master's thesis. Universidade Catolica Portuguesa (Portugal).

[5] Castelnovo, A., 2024. Towards responsible ai in banking: Addressing bias for fair decision-making. arXiv preprint arXiv:2401.08691 .

[6] Černevičiene, J., Kabašinskas, A., 2024. Explainable artificial intelligence (xai) in finance: a systematic literature review. Artificial Intelligence Review 57, 216.

[7] Chernysh, O., Smishko, O., Koverninska, Y., Prokopenko, M., Pistunov, I., 2024. The role of artificial intelligence in financial analysis and forecasting: Using data and algorithms. Economic Affairs 69, 1493–1506.

[8] Choi, I., Kim, W.C., 2023. Enhancing financial literacy in south korea: Integrating ai and data visualization to understand financial instruments' interdependencies. Societal Impacts 1, 100024.

[9] D'Acunto, F., Rossi, A.G., 2023. It meets finance: financial decision-making in the digital era, in: Handbook of financial decision making. Edward Elgar Publishing, pp. 336–354.

[10] Ghosh, I., Jana, R.K., Sanyal, M.K., 2019. Analysis of temporal pattern, causal interaction and predictive modeling of financial markets using nonlinear dynamics, econometric models and machine learning algorithms. Applied Soft Computing 82, 105553.

[11] KARACAN, S., KIRDAR, M., 2021. Hisse senedi fiyat tahmininde makine öğrenmesi ve yapay zeka kullanimi. Journal of International Social Research 14.

[12] Kılıç, A., Güloğlu, B., Yalçın, A., Üstündağ, A., 2023. Big data–enabled sign prediction for borsa istanbul intraday equity prices. Borsa Istanbul Review 23, S38–S52.

[13] Lai, G., Chang, W.C., Yang, Y., Liu, H., 2018. Modeling long-and short-term temporal patterns with deep neural networks, in: The 41st international ACM SIGIR conference on research & development in information retrieval, pp. 95–104.

[14] Lee, J., Kim, R., Koh, Y., Kang, J., 2019. Global stock market prediction based on stock chart images using deep q-network. IEEE Access 7, 167260–167277.

[15] Leung, M.F., Jawaid, A., Ip, S.W., Kwok, C.H., Yan, S., Leung, M., Jawaid, A., Ip, S., Kwok, C., Yan, S., 2023. A portfolio recommendation system based on machine learning and big data analytics. Data Science in Finance and Economics 3, 152–165.

[16] Lundberg, S.M., Lee, S.I., 2017. A unified approach to interpreting model predictions. Advances in neural information processing systems 30.

[17] Mahmudov, A., 2024. Application of machine learning and data analysis in enhancing financial literacy. Bulletin news in New Science Society International Scientific Journal 1, 49–57.

[18] Mukherjee, S., Sadhukhan, B., Sarkar, N., Roy, D., De, S., 2023. Stock market prediction using deep learning algorithms. CAAI Transactions on Intelligence Technology 8, 82–94.

[19] Murugesan, R., Manohar, V., 2019. Ai in financial sector–a driver to financial literacy. Shanlax International Journal of Commerce 7, 66–70.

[20] Niazi, M.K.S., Malik, Q.A., 2019. Financial attitude and investment decision making-moderating role of financial literacy. NUML International Journal of Business & Management 14, 102–115.

[21] Owolabi, O.S., Uche, P.C., Adeniken, N.T., Ihejirika, C., Islam, R.B., Chhetri, B.J.T., Jung, B., 2024. Ethical implication of artificial intelligence (ai) adoption in financial decision making. Comput. Inf. Sci 17, 49–56.

[22] Qatawneh, A.M., Lutfi, A., Al Barrak, T., 2024. Effect of artificial intelligence (ai) on financial decision-making: Mediating role of financial technologies (fin-tech). HighTech and Innovation Journal 5, 759–773.

[23] Rahman, M.H., Tipu, M., Ahmed Habib, K., Ahmed, N., 2023. Development of financial literacy through a self-advising stock investment model: an integration of machine learning and business intelligence tools to guide novice investors. Available at SSRN 4669952 .

[24] Reddy, V.K.S., Sai, K., 2018. Stock market prediction using machine learning. International Research Journal of Engineering and Technology (IRJET) 5, 1033–1035.

[25] Ribeiro, M.T., Singh, S., Guestrin, C., 2016. " why should i trust you?" explaining the predictions of any classifier, in: Proceedings of the 22nd ACM SIGKDD international conference on knowledge discovery and data mining, pp. 1135–1144.

[26] Şişmanoğlu, G., Koçer, F., Önde, M.A., Sahingoz, O.K., 2020. Derin öğrenme yöntemleri ile borsada fiyat tahmini. Bitlis Eren Üniversitesi Fen Bilimleri Dergisi 9, 434–445.

[27] Thakkar, A., Chaudhari, K., 2021. Fusion in stock market prediction: A decade survey on the necessity, recent developments, and potential

future directions. Information Fusion 65, 95–107.

[28] Vaswani, A., Shazeer, N., Parmar, N., Uszkoreit, J., Jones, L., Gomez, A.N., Kaiser, Ł., Polosukhin, I., 2017. Attention is all you need. Advances in neural information processing systems 30.

[29] Vijh, M., Chandola, D., Tikkiwal, V.A., Kumar, A., 2020. Stock closing price prediction using machine learning techniques. Procedia computer science 167, 599–606.

[30] Wawer, M., Chudziak, J., 2025. Integrating traditional technical analysis with ai: A multi-agent llm-based approach to stock market forecasting, in: 17th International Conference on Agents and Artificial Intelligence, SciTePress.

[31] Weber, P., Carl, K.V., Hinz, O., 2024. Applications of explainable artificial intelligence in finance—a systematic review of finance, information systems, and computer science literature. Management Review Quarterly 74, 867–907.

[32] Wu, B., 2023. Is gpt4 a good trader? arXiv preprint arXiv:2309.10982 .

[33] Yılmaz, C., Öztürk, S., 2023. Davranişsal finans ve kamuyu aydınlatma perspektifinde şirketlere ilişkin bildirimler ve haberler: Literatür değerlendirmesi. Muhasebe ve Finans İncelemeleri Dergisi 6, 132–159.

[34] Zeng, A., Chen, M., Zhang, L., Xu, Q., 2023. Are transformers effective for time series forecasting?, in: Proceedings of the AAAI conference on artificial intelligence, pp. 11121–11128.

[35] Zerveas, G., Jayaraman, S., Patel, D., Bhamidipaty, A., Eickhoff, C., 2021. A transformer-based framework for multivariate time series representation learning, in: Proceedings of the 27th ACM SIGKDD conference on knowledge discovery & data mining, pp. 2114–2124.